# Tackling the Algorithmic Control Crisis
# –the Technical, Legal, and Ethical Challenges
# of Research into Algorithmic Agents

Bodo, B.,[1] Helberger N.,[2] Irion, K.,[3] Zuiderveen Borgesius F.,[4] Moller, J., [5] van der Velde, B., [6] Bol, N.,[7] van Es, B.,[8] de Vreese, C.[9]

*Algorithmic agents permeate every instant of our online existence. Based on our digital profiles built from the*

---

[1]    Institute for Information Law, University of Amsterdam. This manuscript was developed as a collaborative effort stemming from a joint research project. The research for this article was made possible with the help of research funding from the University of Amsterdam (Personalised Communication), a research grant from the European Research Council, ERC, under Grant 638514 (PersoNews), and a grant from the SIDN Fonds. The first author created the initial design of, and coordinated the development of the Robin tool described in this article, took the lead in developing the paper, and coordinated writing and integration with the remaining authors. Contact: bodo@uva.nl.

       Beyond the authors of this paper the Robin tool was developed and implemented by a larger team of people, which included: Dr. Damian Trilling, and Mats Willemsen from the University of Amsterdam; Marika de Bruijne, and Arnaud Wijnant (CentERdata); Lieke Beelen and Simon Jimenez (UX design of the plugin and the informed consent process); and Steven Eardley, Mark MacGillivray, Anusha Ranganathan, and Martyn Whitwell for CottageLabs LLP who were responsible for software development and operations.

[2]    Institute for Information Law, University of Amsterdam. Prof. Dr. Natali Helberger contributed to the legal and ethical analysis, the conclusions, and developed the legal, ethical and organizational safeguards around Robin.

[3]    Institute for Information Law, University of Amsterdam. Dr. Kristina Irion contributed to the legal and ethical analysis, the conclusions, and developed the legal, ethical and organizational safeguards around Robin.

[4]    Institute for Information Law, University of Amsterdam. Dr. Frederik Zuiderveen Borgesius contributed to the legal and ethical analysis, the conclusions, and developed the legal and ethical safeguards around Robin.

[5]    Amsterdam School of Communication Research, University of Amsterdam. Dr. Judith Moller contributed to the ethical analysis, and to the definition of goals and methods of the research.

[6]    Information Language Processing Systems, University of Amsterdam. Dr. Bob van de Velde contributed to the coordination, development and testing of the Robin software environment.

[7]    Amsterdam School of Communication Research, University of Amsterdam. Dr. Nadine Bol contributed to the development and testing of the UX and the informed consent process.

[8]    Institute for Information Law, University of Amsterdam. Dr. Bram van Es contributed to the development of the Robin software environment, and to the data analysis toolset.

[9]    Amsterdam School of Communication Research, University of Amsterdam. Prof. Dr. Claes de Vreese contributed to the definition of goals and methods the paper and the Robin tool from a communication science perspective.

*massive surveillance of our digital existence, algorithmic agents rank search results, filter our emails, hide and show news items on social networks feeds, try to guess what products we might buy next for ourselves and for others, what movies we want to watch, and when we might be pregnant. Algorithmic agents select, filter, and recommend products, information, and people; they increasingly customize our physical environments, including the temperature and the mood. Increasingly, algorithmic agents don't just select from the range of human created alternatives, but also they create. Burgeoning algorithmic agents are capable of providing us with content made just for us, and engage with us through one-of-a-kind, personalized interactions. Studying these algorithmic agents presents a host of methodological, ethical, and logistical challenges.*

*The objectives of our paper are two-fold. The first aim is to describe one possible approach to researching the individual and societal effects of algorithmic recommenders, and to share our experiences with the academic community. The second is to contribute to a more fundamental discussion about the ethical and legal issues of "tracking the trackers", as well as the costs and trade-offs involved. Our paper will contribute to the discussion on the relative merits, costs and benefits of different approaches to ethically and legally sound research on algorithmic governance. We will argue that besides shedding light on how users interact with algorithmic agents, we also need to be able to understand how different methods of monitoring our algorithmically controlled digital environments compare to each other in terms of costs and benefits. We conclude our article with a number of concrete suggestions for how to address the practical, ethical and legal challenges of researching algorithms and their effects on users and society.*

**TABLE OF CONTENTS**

## I.        INTRODUCTION

Algorithmic agents permeate every instant of our online existence. As Artificial Intelligence research makes steady advances, as sensors proliferate, and more and more data are being accumulated and shared on data markets, the effectiveness of algorithmic recommendations grows while the costs of personalization drop. Consequently, many wonder if there will still be a space in the future where we remain insulated from direct or indirect exposure to algorithmic agents. In the age of ubiquitous algorithmic agents, will there be still spaces where we are not subjected to A/B tests, tailored advertising, price discrimination, and content recommendations? Will there be spaces in the future which are not controlled, one way or another, by algorithmic agents, and where technology is a neutral arbiter of rather than an active agent in our interactions in and with our environment?

Technology, as always, is deployed before society had the opportunity to come to terms with it. The lack of insight leads to a sense of lost control, drawing anxious responses.[10] Many see algorithmic agents as black boxes,[11] or rather, as black holes, which utilize all available information and grow ever powerful, but still remain invisible to human perception. Just like astrophysicists, scholars of algorithmic agents try to evaluate circumstantial evidence to understand how algorithmic agents operate, but unlike natural scientists, the researchers who study the sociological, political, economic, anthropological, ethical, and legal aspects of algorithmic black boxes regard their object of study as anything but natural or value-neutral. Algorithmic agents, just like any other technology, are embedded in the existing economic, social, and political conditions. They reflect our implicit and explicit hopes and fears, ambitions and shortcomings, and the social conditions in which they are created and used.[12] This means that, despite many techno-optimistic accounts,[13] the discourse on algorithms also reflects the fears and speculations on the adverse effects of algorithmic agents. Strong arguments support the position that algorithmic agents that operate without proper, or flawed, human oversight; or absent of well-defined governance and ethical frameworks, may have negative effects on greater societal norms and values such as the holy triumvirate of *liberté, egalité, fraternité*—or to put it in the language of the existing legal frameworks, fundamental human rights and freedoms, equality, and social cohesion.[14]

---

[10]    *See*, e.g., CASS R. SUNSTEIN, INFOTOPIA: HOW MANY MINDS PRODUCE KNOWLEDGE (2006); ELI PARISER, THE FILTER BUBBLE: WHAT THE INTERNET IS HIDING FROM YOU (2011); VAIRA VIKE-FREIBERGA ET AL., A FREE AND PLURALISTIC MEDIA TO SUSTAIN EUROPEAN DEMOCRACY: THE REPORT OF THE HIGH LEVEL GROUP ON MEDIA FREEDOM AND PLURALISM (2013); Malte Ziewitz, *Governing Algorithms: Myth, Mess, and Methods*, 41 SCI., TECH., & HUMAN VALUES 3 (2015); Frederik J. Borgesius Zuiderveen et al., *Should we worry about filter bubbles? An interdisciplinary inquiry into self-selected and pre-selected personalised communication*, 5 INTERNET POLICY REV. (2016).

[11]    Lucas D Introna, *Algorithms, Governance, and Governmentality On Governing Academic Writing*, 41 SCI., TECH., & HUMAN VALUES 17 (2015).

[12]    To illustrate this point, it worth remembering the controversies around Google seemingly serving racist search results and ads. See, for example, Latanya Sweeney, *Discrimination in Online Ad Delivery*, 11 ACM QUEUE 10 (2013) for discriminative ads served with queries for black sounding names. Despite the popular press's descriptions of racist algorithms, Google's algorithmic agents

Responding to these background conditions, the University of Amsterdam launched a research program to study the effects of, and the normative considerations around, online personalized services in the domains of news, politics, commerce, and health communication.[15] This project has four major aims. First, we hope to identify how algorithmic agents tailor news, political communications, commercial offerings, and health-related information. Second, we want to understand what happens in the personalized and private information cocoons: what information individuals are exposed to and how they interact with algorithmic agents and their recommendations. Third, through the synthesis of the observations of these domains, we attempt to understand what societal effects result from this personalization -- including fundamental social and political changes.[16]  Finally, we seek to assess the results of our empirical inquiries from a legal and normative perspective to

---

(or rather, its developers) hardly engage in deliberate racist behavior. Rather, algorithms learn and reflect the values of their users.

[13]  *See, e.g.,* NICHOLAS NEGROPONTE, BEING DIGITAL (1995).

[14]  Tal Zarsky, *The Trouble with Algorithmic Decisions: An Analytic Road Map to Examine Efficiency and Fairness in Automated and Opaque Decision Making*, 41 SCI., TECH., & HUMAN VALUES 118 (2016).

[15]  The Personalized Communications project is a joint, multidisciplinary initiative between the Institute for Information Law (IViR) and the Amsterdam School of Communication Research (ASCoR) at the University of Amsterdam. As the website states: "The objective of the Personalized Communications initiative is to conduct empirical and normative research on the uses, effects, and implications of personalized communication in the areas of politics, health, and commerce." *The Project*, PERSONALIZED COMMUNICATION, https://perma.cc/DCC5-9ML2 (last visited Dec. 3, 2016). This project focuses on investigating "the uses and implications of personalized communication and information" for individuals, society, and information law and policy. *Id.* The program takes a normative-empirical perspective, creating an environment and infrastructure for normative-empirical research and establishing itself as a central knowledge hub for research in this domain.

[16]  High profile political events in 2016, such as Brexit, and the controversies around the US presidential elections led to serious debates about the effect of algorithmic recommenders on the access to and diversity of news; and the fragmentation and/or polarization of public opinion. In the domain of commerce, algorithmic agents may also unjustifiably discriminate certain consumer groups.

identify arguments and tools for possible policy interventions, assuming the normative assessment provides adequate justification for interference.[17]

Many would agree that these are pressing issues. But how does one go about designing a research methodology for such a project? Operationalizing this undertaking is no small task and "tracking the trackers" poses a range of technical, legal, and ethical challenges and trade-offs. In this article, we hope to share our experiences, challenges, and possibly solutions regarding the challenges of studying algorithmic agents in digital communications.

The article is structured as follows. In Part II we argue that our societies need to think seriously about how to do research on algorithmic agents (or "AAs"). We describe the stakes of this research and assess if our current approaches are commensurable with those stakes. We then spell out what it would take to stay in control of our AA-controlled digital environments. In Part III, we give an overview of the current AA research landscape and describe the technical, legal, ethical, and practical difficulties associated with the different research approaches. We also point out some of the fundamental ethical and legal issues around researching algorithmic agents. In Part IV, we describe a novel approach to AA research, which we devised. We also detail how we addressed the previously-mentioned technical, legal, and ethical design challenges. In the conclusion, we summarize our experiences and lay the groundwork for future research and actions. We argue that besides shedding light on the internal workings of specific algorithms, we also need to be able to understand how different methods of AA research compare to each other in terms of costs and benefits.

## II.    THE NEED FOR RESEARCH ON ALGORITHMIC AGENTS

Algorithmic agents have unprecedented control over multiple aspects of our society and lives. This is primarily due to three recent developments in the domain of digital, networked communications. First, it seems that contrary to all the wishful thinking,[18] the currently prevailing technological, commercial, and political incentives are more likely to lead to a heavily centralized    communication    infrastructure    than    to    a decentralized one.[19] In the last few years, we have witnessed the emergence of a small number of extremely powerful intermediaries - such as Facebook in social networking; Google in searching, advertising, and mobile communications; Amazon in online retail and cloud services; and Apple in mobile hardware and software – which have managed to secure their dominant positions in multiple global marketplaces.

---

[17]    *Id.*
[18]    *See, e.g.*, YOCHAI BENKLER, THE WEALTH OF NETWORKS: HOW SOCIAL PRODUCTION TRANSFORMS MARKETS AND FREEDOM (2006).

The second trend is the increased reliance on algorithmic software agents by intermediaries, big and small, to serve and interact with their users and customers and to provide personalized services for them. Google's PageRank algorithm compiles our search results and manages our emails. Facebook's newsfeed algorithm compiles our daily dose of news, friend updates, and cat videos, effectively controlling both our news diet and our social relations. Online shops recommend us goods and try to guess our maximum willingness to pay. Recent advances in Artificial Intelligence research and applications suggest that we should expect algorithmic agents to be applied to even more, currently human controlled domains.

Third, due to the proliferation of algorithmic personalization, an increasing fraction of our digital experience is unique to us, and unknown to others, isolating our digital experiences into individual experience cocoons.

Taken together, these developments present us with pressing challenges about how to stay in control of digital environments which are increasingly co-habited and controlled by opaque algorithmic agents, weaving nontransparent personalized-experience-cocoons around individual users. This non-transparency creates what we call an "algorithmic control crisis", in which we need to solve multiple, deeply intertwined problems that tend to lack well-tested theoretical, methodological, and practical ways to address them.

This "control crisis" is the result of many factors. First, our empirical knowledge of our algorithmically personalized digital environment is fragmented and unmethodical: we lack systematic insight into what is happening inside the individual experience cocoons and how those events aggregate on a societal level. Second, we the lack established benchmarks and thresholds by which to measure and assess changes on the individual and societal level, such as the diversity of information-diets or the fragmentation of public discourse. Third, we lack systematic research into the normative implications of algorithmic control. Fourth and finally, we lack research on the effectiveness of current legal and policy tools to control algorithmic agents, the contexts in which they are deployed, and the interactions they are engaged in.

---

[19]   Jean-Christophe Plantin et al., *Infrastructure studies meet platform studies in the age of Google and Facebook*, NEW MEDIA & SOC. 1 (2016), https://perma.cc/CGV9-55L9.

The scope of this paper is limited. We intend to reflect upon the first challenge, and the first challenge alone: on the fragmentation of our knowledge of algorithmically controlled, personalized information environments. This fragmented knowledge landscape poses particular challenges to researchers attempting to study the effects algorithmic agents.

Our individual experience of algorithmically personalized services is by definition *unique* and differs from everyone else's experiences by an unknown degree. Our experiences are also *private*: in most cases, individual users are alone in those situations where they are exposed to personalized services.[20] Taken together, this means that non-transparent algorithmic personalization agents create non-transparent, unique experience cocoons that remain unknown – perhaps even unknowable - and thus incommensurable for everyone else. We can neither take for granted that our neighbors have the same exposure to the world as we do, nor be sure what others are exposed to.

This creates an unprecedented level of information asymmetry for the individual and for society. The individual is losing the guidance that the evening TV news, the front page of the daily newspaper, the campaign poster, or the shop-window on the high street provided on others' experiences. Society, on the other hand, has yet to develop a capacity to monitor what is happening inside the opaque algorithms, or in the fragmented experience bubbles, and aggregate them into a meaningful whole. The same non-transparency that prevents the individual from trusting the existence of a shared experience with others prevents society as a whole from creating an aggregate view of what is happening in those personalized experience cocoons. The result is that we lose sight of what is happening to us in the era of personalized algorithmic recommendations.

This lost perspective undermines the ability to self-reflect -- and to change course if necessary. Policy, regulation, and control hinge on our ability to monitor information. There is, of course, always an inherent information asymmetry between the regulator and the market, but how big that asymmetry is, and how difficult it is to bridge, varies. Take for example traditional media markets, which seem to have been very transparent compared to what we have today. Information on the supply side: on media products, on their owners, on their employees, on their circulation, on their audience, on their advertisers, and on their potential income was more or less in the public domain, or very cheap to produce. Information gathering at the demand side, at the consumers level, was not much more difficult, as private companies (like Nielsen, or GfK) and public institutions (like the National Bureaus of Statistics) have been conducting systematic, longitudinal, and representative studies on media consumption. However, our societies lack anything even remotely similar for the digital, personalized media. When it comes to the aforementioned digital intermediaries, we as a society have no idea what information and ads individuals are exposed to: we have no way of knowing how that information was selected for them; we do not know whether there is a human editor who edits information streams, and if there is, who he/she might be; and even producers, whose content is being relayed, have only very limited information on who their audience is, while the public has almost no insight into the transactions and information flows on these platforms. The incentives are structured so that whatever limited information stakeholders have on the personalized, digital media market, the information will not be shared, so any meta-information on these markets remains extremely fragmented state, if in any state at all.

---

[20]   Personalization is usually based on data collected on the individual user, aggregated into individual user profiles. There are some probable exceptions to that, such as a Netflix profile, which aggregates preferences on a household level, but shared consumption of personalized services is the exception to the rule of the atomized consumer.

Personalized digital media is not the first market to suffer from such structural information asymmetry: banking, for example, is also historically and structurally non-transparent. For personalized digital media, however, in the foreseeable future we will continue to lack the tools that emerged to provide public oversight of the financial system: statutory transparency obligations, stress tests, and independent audits. Yet, given the power of these digital intermediaries, there is an immediate and pressing need for information so we have a chance to understand what is happening to us, our neighbors, and our societies in the rapidly evolving, increasingly dominant, AA-assisted, personalized digital media environment.

The central position of non-transparent algorithmic agents in our media environments creates the urgent need for research into these agents and their effects. This need has prompted a fresh wave of research into the methods of auditing those algorithmic agents. In the next section, we describe the current landscape of AA audit approaches.

### III.    THE STATE OF RESEARCH INTO ALGORITHMIC AGENTS: THE AUDIT APPROACH IN PERSONALIZED MEDIA RESEARCH, AND BEYOND

#### A.    *The Algorithm Audit Approach*

The initial reaction to the emergence of algorithmic agents was to demand algorithmic transparency via the audit of algorithms. Several different algorithm audit approaches were proposed[21] to shed light to the *inner workings* of non-transparent and complex algorithmic agents:

*Audit of disclosed code (Algorithm Transparency).*[22] This approach operates under the assumption that the public or select individuals have full access to the source code of the algorithmic agent, and thus it is possible to review what kinds of input variables it uses and how these inputs are being used to produce its output. Subsequent studies identified many limitations of this approach: algorithms cannot be separated from the datasets they use and the applications they are used for; discriminatory algorithmic decisions are hardly hard-coded, and may be the emergent properties of the machine learning process, not identifiable from the review of code; full code transparency may actually aid the abuse of the algorithms by malevolent agents, and so on. In any case, there are very few algorithmic agents whose full code is available for review either by the public or by a closed group of experts. Even the most transparent companies in this domain (such as Reddit) keep parts of their code closed to avoid abuse.[23]

Not having direct access to the algorithmic agent itself, the other alternative is to study its outputs, while, if possible, strategically manipulating its inputs to *reverse engineer* how different outputs depend on the changes of inputs. *Scraping audits* use the public interfaces, or APIs, of algorithmic agents to feed them particular information and analyze the outputs. However, in most cases, such APIs are non-existent or have severe limitations. In those cases researchers conduct "*sock*

---

[21]    The two most important contributions, on which the following section is based, are Christian Sandvig et al., *Auditing Algorithms: Research Methods for Detecting Discrimination on Internet Platforms*, Paper Presented to "Data and Discrimination: Converting Critical Concerns into Productive Inquiry," A Pre-Confrence at the 64th Annual Meeting of the International Communication Association (May 22, 2014), https://perma.cc/V5T8-WQKC; and Rob Kitchin, *Thinking Critically About and Researching Algorithms*, The Programmable City Working Paper 5 *1 (Oct. 28, 2014).

[22]    Sandvig, *supra* note 21, at 9-10.

[23]    Randall Munroe, *Reddit's new comment sorting system*, REDDIT BLOG (Oct. 15, 2009),  https://perma.cc/N3LT-WKU3.

*puppet audits*." This method entails often sophisticated software and hardware infrastructures[24] designed to emulate certain aspects of user behavior in relation to algorithmic agents. Such infrastructures are usually able to control variables researchers think may be relevant inputs for the algorithmic agents, such as usage histories, technical parameters (browser and operating system type), geolocation, etc. Though recent years brought significant improvements in the sophistication of the audit infrastructure,[25] simulated users may or may not be adequate approximations of real-life users. While such puppet audits may yield noteworthy results of how artificial changes in the synthetic profiles result in changes in the algorithmic agents output, without proper benchmark against real-life users, it is impossible to say whether those observed effects are generalizable beyond those artificially created contexts.

The solution to the shortcomings of sock puppet audits is to observe how algorithmic agents react to the inputs of real-life users. In *Crowdsourced / Collaborative Audits*, real user-algorithmic agent interactions are observed, and the pre-existing profiles, browsing histories, technology fingerprints, and other organically developed profile information are used. However, this proximity to reality comes at a cost. First, in real-world observation, it is often problematic to control for independent variables and differences in conditions. In addition, recruiting a large enough sample that is representative and diverse in those dimensions in which algorithmic discrimination or unjust profiling might be a relevant issue is costly and difficult. Finally, and maybe most challenging, studying algorithms in their relationship to "real" humans raises a host of legal and ethical issues, which we will describe in more detail in the next section.

Finally, a number of *qualitative approaches* can help researchers uncover the context in which algorithmic agents are designed, deployed, and interacted with. The full socio-technical assemblage around algorithmic agents,[26] or "networked information algorithms,"[27] interpret algorithms in their wider contexts of the conditions of code development and deployment, usage, interfaces, and data, as well as the normative, legal, organizational, economic, political, and cultural frameworks in which code is situated and may be interpreted. Unpacking the full sociotechnical assemblages of algorithms may be difficult to achieve due to the complexity of these assemblages, but partial reconstructions may be possible through the (auto)ethnographic studies of their genesis or the observation of algorithmic agents' operation in their native contexts. When engaged in *reflexively producing code,* researchers "reflect[] on and critically interrogate[] their own experiences of translating and formulating an algorithm."[28] Such studies provide insight into the legal, ethical, institutional, socio-technical contexts in which code development needs to be situated. Rather than conducting auto-ethnography, it is also possible to conduct *ethnographic studies on the coding teams.* Examining the professional teams deploying algorithms may reveal how different decisions shape the ways algorithmic agents are developed, customized to specific tasks, or deployed in light of different technological, editorial, and business considerations.[29]

---

[24]   For a description of such an infrastructure, see Steven Englehardt et al., *OpenWPM: An automated platform for web privacy measurement* (Mar. 15, 2015), https://perma.cc/S2G4-F74H.

[25]   *Id.*

[26]   Kitchin, *supra* note 21, at *7-13, *21.

[27]   Mike Ananny, *Toward an Ethics of Algorithms Convening, Observation, Probability, and Timeliness*, 41 SCI., TECH. & HUMAN VALUES 93 (2015).

### B.        *Beyond the Audit of Algorithms: Inquiry Into the User*

Most, if not all, of the currently dominant, aforementioned research strategies focus on the algorithm itself. They all rest of the premise that one way or another it is sensible, possible, and effective to model how algorithmic recommendations are *produced*, so they can be subjected to what is essentially a supply-side analysis. It is hoped that this way algorithmic agents can be subjected to an *a priori* (deontological) ethical scrutiny (do they have the appropriate ethical guidelines encoded?), or to a teleological critique to test whether these agents produce legally and ethically acceptable results.[30] As we have seen, even with a fully transparent code, the inner workings of an algorithmic agent may remain unintelligible for humans, making the *a priori* scrutiny hard, if not impossible. The consequentialist approach would require us to model all possible users and all possible circumstances to account for all the possibilities that might arise. This also seems rather difficult to achieve. In addition, in both cases the results would remain detached from what is actually happening to flesh and blood humans under very specific and real conditions.

The alternative to the study of algorithms is the study of the effects of algorithmic agents on individual users. Individual users engage with algorithmic agents every day; they use and abuse, cheat, resist, play with and subvert what algorithms are

---

and offer. The personalized experience cocoons around each individual are the products of these interactions between humans and algorithmic agents. Contrary to the algorithm-audit approach, personalized experience cocoons are not specific to any particular algorithmic agent, but they reflect the whole spectrum of online and offline, personalized and non-personalized information flows.

To illustrate this point, take, for example, news personalization. Despite the recent growth in news personalization, personalized sources likely constitute only a small share of a person's news diet, especially if we factor in other news media, such as television, radio and print. On the other hand, heavily personalized services, such as social networks and search engines, are significant traffic drivers for personalized news sites. Any research that hopes to reconstruct the effects of news personalization, must be able to observe all of the distinct, but closely interrelated, personalized and non-personalized online and offline domains that account for the news diet of the individual, and shape individuals' implicit and explicit personalization choices.

For these reasons we argue that rather than looking at algorithmic agents in isolation, we need to focus on the co-development of non-personalized media, algorithmic personalization agents, and users. The individuals who interact with algorithmic agents - who rely on, ignore, or resist personalized recommendations - are not the passive victims of algorithmic agents, but key stakeholders, with full agency. These interactions both shape the algorithmic agents, and produce information on their workings. Only through the aggregation of these individual observations can one fully view the actual individual and societal benefits and harms of algorithmic personalization.

This requires the direct, systematic, automated observation of the online activities of internet users. Not having access to the information that is being gathered at the intermediaries, monitoring the digital information environment of a representative population sample is the second best - and perhaps only - way to reconstruct the individual and societal effects of algorithmic interference in our information flows.

This approach, however, creates an almost unresolvable paradox. Studying the processes of dataveillance,[31] digital profiling, and the algorithmic control of information and users is impossible without interfering with the privacy of the individuals under monitoring. Even if such monitoring is limited, voluntary, and complies with the highest legal and ethical standards, it would capture massive amounts of personal data. Needless to say, this raises serious privacy, data protection, and ethical issues. In other words, those who wish to study algorithmic agents have to reflect on an unavoidable challenge: we can only achieve the benefits of understanding the societal effects of algorithmic agents by surveilling individual users, and thus interfering with individual rights and liberties.

---

[31] *See* Roger Clarke, *Information Technology and Dataveillance*, 31 COMMS. ACM 498 (1988).

## IV.    THE CHALLENGES OF DESIGNING AN ALTERNATIVE: ROBIN – OUR MONITORING TOOL FOR ALGORITHMICALLY PERSONALIZED DIGITAL MEDIA

In this Part, we present an overview of the challenges to balancing these individual and societal interests. We do so through a self-reflexive account of developing our own alternative to the approaches described above: "Robin," a custom-built browser plug-in designed to collect information about the effects of algorithmic agents.

We present this as a type of auto-ethnography, which attempts to unpack the full socio-technical assemblage around our own attempts to observe the algorithmic society, and around the scientific, technical, ethical, and organizational considerations we took during the development process.

We first present the design of the technology to monitor individual information cocoons – that is, how Robin actually *works*. We then address the ethical, legal and organizational challenges and concerns related to this approach.

### A.    *The Technical Design of Robin*

In order to study what happens in personalized information cocoons, we need to capture all relevant types of information exchanged between a consenting subject's internet browser and all online services that might play a role in dataveillance, profiling, and personalization. For that reason, we developed a software tool to intercept data traffic between the browser and the internet. The tool, Robin, is a custom-built browser plugin, which routes the data stream generated by the internet browser of the user through an enhanced proxy server[32] where the data stream is copied, filtered, and stored. This setup (which resembles a Man-In-The-Middle setup – though of course without the malicious intent and the subsequent theft of personal data) enables us to observe either directly or indirectly all the elements of personalization and algorithmic recommendation. First, we are able to capture every piece of information that a user may knowingly or unknowingly expose to online services via various online trackers, beacons, cookies, hardware and software fingerprints, IP addresses, etc. Second, we are able to see which pieces of information the individual is exposed to: what news items he or she sees; what his or her search queries produce; what ads he or she is being served; what prices he or she receives, etc. Finally, we are able to capture user interaction in terms of comments, likes, shares, follow up-searches, etc.

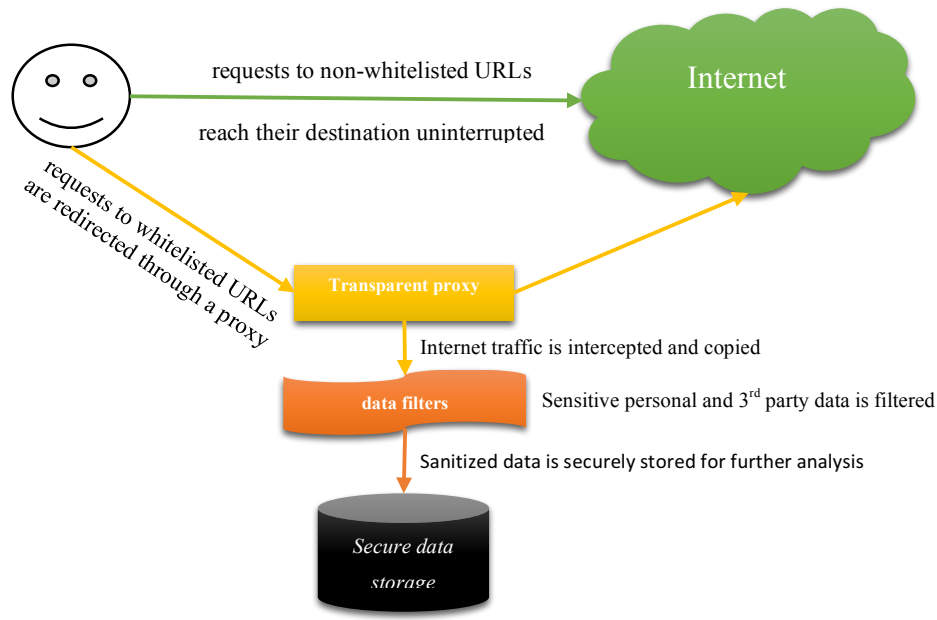

***Figure 1: Basic schemata of the observation infrastructure***

The monitoring relies on a crowdsourced / collaborative approach: we planned to recruit 1600 participants from a well-

---

established social science survey panel[33] to install the plugin. This way we could rely on the existing browsing histories and, if authenticated, pre-existing online profiles of participating users, and we can also survey non-visible characteristics such as participant attitudes.

This relatively simple technical design allows us to capture all data algorithmic agents use and produce. The difficult part was to define the data we *did not* want to collect; to define, circumscribe, and filter out sensitive, unnecessary and/or private data from the captured data stream. On the one hand, we need enough data to be able to reconstruct what happens in those isolated experience cocoons. But on the other, this has to be done in a manner compliant with European legal requirements and consistent with the high ethical standards of the project. The real challenge is thus how to balance data collection with the protection of respondents' privacy, personal data, and security. These are addressed below.

## B.     Robin:     The     Challenge     of     Design     and Implementation

Our drive towards collecting more data is moderated by several additional external and internal factors. First, we are constrained by external *data needs* regarding what information we sought to collect. Built-in *technological roadblocks* limit what is physically possible to collect. The existing *legal frameworks* around, for instance, privacy, data protection, and copyright define the rigid external limits of what can legally be collected. The *panel constraints* include limitations set by the research company, which is responsible for the prolonged existence of their research panel, and the individual sensibilities of the panel respondents, who can reject participation if they find the terms unacceptable or the compensation insufficient. Finally, our formal (as embodied in institutional ethical review boards) and informal *ethical considerations* shape our research.

### 1.     *Data needs when studying personalized communication*

---

[33] Centerdata administered the LISS panel, a publicly funded research panel set up to enable social science studies in the Netherlands. This is similar to the US-based Knowledge Networks. The study used the LISS panel. The role of Centerdata is to assemble a sample of users who are willing to install our browser plugin; inform users about the research, obtain their informed and explicit consent, and manage the panel throughout the 24 months of data collection period.

There are two considerations that define the breadth and scope of data required to reliably consider the normative implications of algorithmic personalization: (1) the spectrum of interactions that may directly or indirectly shape algorithmic personalization; (2) the constitution of the observed group that allow for the generalization of findings.

The study of how people interact with algorithmic agents must rely on two distinct types of information. First, we need to understand what is happening to the individual users within the algorithmically personalized experience cocoons: what kinds of data are collected about them, what kind of profiles are being built, how those profiles are translated into actual algorithmic decisions, and what kind of interactions follow those algorithmic decisions. Second, to piece together the whole picture of the digital public sphere, researchers must also have information on user practices, interactions, activities which do not directly involve algorithmic agents, but are relevant in the context of, for example, news consumption. Consequently, researchers who intend to understand how profiling and targeting works, and whether it leads to (unintentional) biases, are faced with the puzzle that in order to draw conclusions, they need to obtain even more data than any particular profiler in isolation. For example, the aforementioned sock-puppet based approach, which is the preferred method of computer scientists to reverse engineer algorithmic decision-making in the context of a particular algorithm, does not enable the reconstruction of the diffusion of news as users fluidly move between personalized and non-personalized domains.

We do not believe that it is possible to achieve this level of oversight with a single tool. For example, to measure potential effects of exposure to personalized news use, and to test if people become more knowledgeable due to personalized news use (because, for instance, the news selected for them perfectly matches their interest and cognitive capabilities), additional data are needed, that must, for example, be collected through an online knowledge test in a quasi-experimental setting. For these reasons we make intensive use of surveys, interviews, and focus group discussions to complement the data collected through Robin.

The second issue concerns the generalizability of findings. Ultimately, the goal of this research is to generate generalizable insights. This means it is important that our findings be translatable to other users in other circumstances. To meet this goal, it is essential that we do not collect data among a small selective group of users, but that we draw a relatively large random sample of the overall population. If we find that the mechanisms we identify are comparable among all segments of the population, we can assume that they are generalizable and similar results can and will be found if the research is carried out again. It also means that the causal mechanisms and patterns we find hold empirical relevance which is significant for the legitimacy of any future legal intervention.

### 2.        *Technological roadblocks*

In theory, we face very few technical restrictions on the data we can collect on the online behavior of our panel participants. By inserting ourselves between the browser and the internet, we are in a position to observe all data sent and received by the browser, including encrypted information, which we decrypt, save, filter and re-encrypt at the proxy. By having access to the browser framework, in theory we can also use the data generated by other components of the computer: capture mouse movements, access data generated by the camera, the microphone, etc.

At the time of publication, our approach covers PCs, but excludes mobile equipment, such as smart phones and tablets, and smart TVs and other appliances. That is our most important limitation, as mobile devices command an ever-increasing share of our online time. The technological constraints of mobile platforms and smart appliances (the general lockdown of devices, lack of browser plugins, the proliferation of apps that are kept strictly isolated from each other to prevent data leakage and protect user privacy, etc.) severely limit the ability of third parties to track what people do on such devices. With properly configured devices,[34] however, the transparent proxy approach has the potential to observe user activity on currently inaccessible devices.

The main technological roadblock is not how to observe certain user interactions, but how to filter out data we do not want to capture. There are two technological instruments to define what is captured and what is not. First, websites that we want to observe are placed on a whitelist which limit the scale and magnitude of the data collection to what we deem necessary in the light of the research questions. Consequently, data is only routed through the transparent proxy if the user visits a whitelisted website.

In addition, all data is filtered for certain categories of personal data which we identified as highly sensitive personal data. For example, passwords, credit card data and other financial information, personal correspondence, and personal data of individuals who did not consent to being observed are among those data that needs to be filtered out in order to comply with legal and ethical requirements. Tailor-made filters, designed to fit the particulars of each observed website are the second set of technological instruments we employ to limit the invasive nature of our research and comply with the legal and ethical restrictions.

---

[34]    The pre-configuration of devices is necessary, for example, to route all or selected traffic through a proxy, and capture encrypted data traffic.

### 3.     Relevant legal and ethical frameworks

Thus, the largest constraints are legal and ethical. This Section introduces the requirements stemming from legislation on personal data protection and explains how these requirements influenced our approach to collecting and using participants' personal data.[35]

### a)     Complying with EU Data Protection Law

The European Union (EU) vests a high level of legal protection to an individual's personal data. Scientific research in the EU and the US acknowledges similar ethical values. But EU-based research must adhere to the EU rules on privacy and personal data – rules that are different than privacy rules in the US.[36] For instance, the notion of "personal data", a legal term in EU law, differs from what would be considered personal identifiable information in US law.[37]

In European constitutional law, the right to private life - commonly referred to as the right to privacy - has the status of a fundamental right under Article 8 of the European Convention on Human Rights[38] and Article 7 of the Charter of Fundamental Rights of the European Union (EU Charter).[39]

---

[35] Research with consenting participants into online personalized services in the Netherlands must comply with relevant legal and ethical frameworks. However, the intended observation of participants' exposure to online personalized services can also potentially conflict with intellectual property law and contract law, which we have kept outside the scope of this paper.

[36] For a review of the US legal environment, see David R. O'Brien et al., *Integrating Approaches to Privacy Across the Research Lifecycle: When Is Information Purely Public?* (The Berkman Center for Internet & Society at Harvard University, Research Publication No. 2015-7, 2015)., https://perma.cc/PY8U-LN96.

[37] Paul M. Schwartz & Daniel J. Solove, The PII Problem: Privacy and a New Concept of Personally Identifiable Information, 86 N.Y.U. L. REV. 1814, 1877–78 (2011); Frederik J. Zuiderveen Borgesius, *Singling Out People Without Knowing Their Names – Behavioural Targeting, Pseudonymous Data, and the New Data Protection Regulation*, 32 COMPUTER LAW & SECURITY REVIEW 256 (2016);

The EU is exceptional in that the EU Charter, entering into force in 2009, grants each person a fundamental right to the protection of personal data concerning him or her. Article 8 provides that personal data "*must be processed fairly for specified purposes and on the basis of the consent of the person concerned or some other legitimate basis laid down by law.*" This right is fortified by granting individuals the right to access their personal data and, if inaccurate, to have the data rectified.

EU data protection law, which aims for complete and effective protection of these fundamental rights, regulates in detail the responsibilities and rights in connection with the handling of personal data. Today, the use of personal data is governed by the 1995 Data Protection Directive,[40] which EU Member States implemented into their national legislation. In the Netherlands, where the research project is based, the implementing law is the Dutch Data Protection Act, which has to be observed when processing personal data.[41]

The legal framework will change in the near future. In May 2018, the EU's General Data Protection Regulation will enter into force and repeal the Directive.[42] As an EU regulation, the General Data Protection Regulation will become directly applicable law in the Member States. The new regulation is bound to change the situation of scientific research because it treats this purpose as desirable for society.

EU data protection law applies almost without exception to personal data used in the course of scientific research.[43] The legal requirements for scientific research involving personal data in EU member states are different from US human subject research regulations.[44] Notably, in EU Member States it would not make a difference if research uses exclusively public information because data protection law applies whenever personal data are used.[45]

---

[38]   Convention for the Protection of Human Rights and Fundamental Freedoms, art. 8, Nov. 4, 1950, 213 U.N.T.S. 221, 230 [hereinafter European Convention on Human Rights or ECHR].

[39]   Charter of Fundamental Rights of the European Union art. 7, 2000 O.J. (C 364) 1.

[40]   Directive 95/46/EC of the European Parliament and of the Council of 24 October 1995 on the Protection of Individuals with Regard to the Processing of Personal Data and on the Free Movement of Such Data 1995 O.J. (L 281) 31[hereinafter Data Protection Directive].

[41]   Wet bescherming persoonsgegevens [Personal Data Protection Act], Stb. 2000, 302 (Neth.).

[42]   *See* Regulation (EU) 2016/679 of the European Parliament and of the Council of 27 April 2016 on the protection of natural persons with regard to the processing of personal data and on the free movement of such data, and repealing Directive 95/46/EC, 2016 O.J. (L 119) 1 [hereinafter General Data Protection Regulation or GDPR].

[43]   The exception for scientific research, which is explained later in the article, provides a legal basis only for the reuse of lawfully collected personal data.

In the following, we map out how, in our effort to comply with local laws, we had to adjust our research methodology and infrastructure. In doing so, we also explain some key concepts of EU data protection law.

Data protection law is triggered whenever 'personal data' are collected or otherwise processed.[46] Personal data are defined as *"any information relating to an identified or identifiable natural person ('data subject'); an identifiable person is one who can be identified, directly or indirectly, in particular by reference to an identification number or to one or more factors specific to his physical, physiological, mental, economic, cultural or social identity."*[47] The new General Data Protection Regulation explicitly mentions "*online identifier*" and "*location data*" as examples of identifiers.[48]

Whether the definition of personal data is met depends on the circumstances of the situation. Merely changing a name to a number is generally not sufficient to render personal data "anonymous" and to remain outside of the scope of data protection law. This wide scope of data protection law is often overlooked in the practice; there is frequent confusion and discussion about the notions of "pseudonymous" and "anonymous" data.[49]

In brief, personal data can relate to a person even if no name is attached, as long as the data are not aggregated to the stage that they cannot be linked anymore with reasonable effort to the individual. Removing the name from a file and replacing it with a number, which then is stored in a separate file, does not in itself make the data being anonymous. Rather, such data have been pseudonymized.[50] And pseudonymized data still fall under EU data protection law, unlike anonymized data.[51] For data to be truly anonymized, the data would need to be altered in a way that it is not any longer possible to trace them back to an identified or identifiable natural person.[52]

---

[44]   *See* 45 C.F.R. § 46.

[45]   *See* Case C-131/12 Google Spain SL v. Agencia Española de Protección de Datos (2014).

[46]   Processing is defined very broadly. *See* Data Protection Directive art. 2(b).

[47]   *Id.* art 2(a).

[48]   *See* General Data Protection Regulation art. 4(1).

[49]   *See* generally about the scope of the personal data definition:  Frederik J. Zuiderveen Borgesius, *Singling Out People Without Knowing Their Names – Behavioural Targeting, Pseudonymous Data, and the New Data Protection Regulation*, 32 COMPUTER LAW & SECURITY REVIEW 256 (2016); *see also* Gerrit-Jan Zwenne, De Verwaterde Privacywet [Diluted Privacy Law], Inaugural lecture of Professor Dr. G. J. Zwenne to the office of Professor of Law and the Information Society at the University of Leiden on Friday (Apr. 12, 2013), https://perma.cc/W7KW-J546..

[50]   *See* General Data Protection Regulation art. 4(5):

Thus, in our research we handle participants' personal data and consequently data protection law fully applies to our research. Moreover, we aim to collect website traffic, revealing participants' news consumption, health-related information, commercial transactions, and exposure to targeted ads. Such website traffic can disclose, or at least suggest, an individual's political opinions, or give information on users' level of fitness and health condition. This implies that we collect "special categories of data". For such "special categories of data", the rules are stricter. In fact, processing such special categories of data is in principle prohibited.

However, this in-principle prohibition can be overridden in a few narrowly defined circumstances, or with the informed and *'explicit'* consent of the research participant (data subject).[53] Scientific research is not mentioned as an exception yet and thus, for the intended observation of participants' exposure to online personalized services, we have to obtain the participant's consent.

In summary, our study clearly falls within the bounds of existing EU law. Thus, we must ensure that we comply with the core data protection principles embodied in the legislation. These are addressed in turn.

i)          Data Minimization

Under the principle of data minimization, personal data must be *"not excessive in relation to the purposes for which they*

---

'pseudonymisation' means the processing of personal data in such a manner that the personal data can no longer be attributed to a specific data subject without the use of additional information, provided that such additional information is kept separately and is subject to technical and organisational measures to ensure that the personal data are not attributed to an identified or identifiable natural person.

[51]     *Id.*

[52]     The Article 29 Working Party has defined anonymous data as: "any information relating to a natural person where the person cannot be identified, whether by the data controller or by any other person, taking account of all the means likely reasonably to be used either by the controller or by any other person to identify that individual." Article 29 Working Party, Opinion 4/2007 on the concept of personal data, (June 20, 2007), *https://perma.cc/L4TR-EDKL*; *see also* Article 29 Working Party, Opinion 05/2014 on anonymisation techniques (Apr. 10, 2014), *https://perma.cc/HZT6-B7EV.*

[53]     Data Protection Directive art. 8(2)(a).

*are collected and/or further processed.*"[54] This data minimization requirement clashes with collecting as much research data as possible. Hence, as researchers we will have to define what personal data we need to collect in relation to a specific research purpose. A bulk data collection approach without a data minimization strategy would breach the data minimization principle. We thus need to develop ways to justify what we collect, which we do in the form of a white list and customized filters to minimize data capture, further described below.

ii)    Storage Limitation

According to the storage limitation principle, personal data must be *"kept in a form which permits identification of data subjects for no longer than is necessary for the purposes for which the data were collected or for which they are further processed."*[55]

From a research perspective, there are actually good reasons to store data indefinitely. For instance, storing data indefinitely makes sense, because we might think of new research questions later. If doubts ever arise about research results, researchers want to be able to provide doubters with the original data set. If applied too narrowly, the legal requirement of storage limitation conflicts with the ethical expectation of data openness for accountability and the reproducibility of results. Our research design takes a rolling approach to the question of how long the pseudonymized dataset will be kept and maintained by stipulating that the unprocessed personal data will be deleted five years after the last publication was published.

iii)    Purpose Limitation

The purpose limitation principle requires that personal data must be *"collected for specified, explicit and legitimate purposes and not further processed in a way incompatible with those purposes."*[56] The purpose limitation principle, a core data protection principle, has several implications for the research project.

First, the purpose limitation principle makes it illegal to release (without a participant's consent) personal data as 'open data', as far as 'open data' implies that anyone can use the data for any purpose.[57] We can, however, disclose aggregated data sets, such as regression tables, as such data sets do not qualify as personal data. But there is much public and scientific interest in 'open data': making datasets collected by scientists (who are tax-funded) available for other scientists.[58] Access to scientific and research data is possible today because the EU legal framework has an exception that allows, under certain circumstances, further data processing for scientific research.[59] However, the General Data Protection Regulation might introduce additional obligations for researchers, in addition to clarifying the safeguards that researchers must put in place when handling personal data.[60]

---

[54]    Data Protection Directive art. 6(1)(c).

[55]    *Id.*, art. 6(1)(e).

[56]    *Id.*, art. 6(1)(b).

[57]    *See generally* Frederik Zuiderveen Borgesius, Jonathan Gray & Mireille van Eechoud, *Open Data, Privacy, and Fair Information Principles: Towards a Balancing Framework*, 30 BERKELEY TECH. L.J. 2073, 2132 (2015).

Second, the purpose limitation principle requires that the purpose of collecting and using personal data be defined in advance. Hence, we need to precisely define the collection purpose, which can be challenging seeing that this a multi-year, multi-project research initiative in which we seek to create a research infrastructure without necessarily knowing all possible research questions in advance. The new European Data Protection Regulation explicitly recognizes that *"[i]t is often not possible to fully identify the purpose of personal data processing for scientific research purposes at the time of data collection.*"[61] Therefore, data subjects should be allowed to give their consent to certain areas of scientific research when in keeping with recognized ethical standards for scientific research. The regulation leaves it to the member states to describe specific research exemptions, providing they arrange necessary safeguards.[62] We defined the purpose of data collection as "research into the effects of personalized communication."

---

58  *See, e.g.*, Data Access & Research Transparency, DA-RT, https://perma.cc/LEH9-JSVB_ (last visited Nov. 30, 2016).

59  *See* Data Protection Directive art. 6(1)(b).

60  General Data Protection Regulation art. 5(1)(b); *see also* art. 89(1).

61  *Id.* recital 33.

62  *Id.* recital 23. "Member States should be authorised to provide, under specific conditions and in the presence of appropriate safeguards for data subjects, specifications and derogations to the information requirements, rectification, erasure, to be forgotten, restriction of processing and on the right to data portability and the right to object when processing personal data for archiving purposes in the public interest, or scientific and historical research purposes or statistical purposes. The conditions and safeguards in question may entail specific procedures for data subjects to exercise those rights if this is appropriate in the light of the purposes sought by the specific processing along with technical and organisational measures aimed at minimising the processing of personal data in pursuance of the proportionality and necessity principles." *See also* art. 83 ("Processing of personal data for archiving purposes in the public interest, or scientific and historical research purposes or statistical purposes, shall be subject to in accordance with this Regulation appropriate safeguards for the rights and freedoms of the data subject. These safeguards shall ensure that technical and organisational measures are in place in particular in order to ensure the respect of the principle of data minimisation. These measures may include pseudonymisat ion, as long as these purposes can be fulfilled in this manner. Whenever these purposes can Finally, the purpose limitation principle requires that personal data be collected for "legitimate purposes." This phrase refers to the requirement in EU data protection law for data controllers (data users) to have a "legal ground" or "legal basis" to process personal data. In principle, there are six legal bases that data controllers can rely on to process personal data. But our research can only rely on the data subject's informed consent.[63] In our project we collect "special categories of data", which, in short, can only be lawfully processed if the data subject has given his or her "explicit consent", or when a specified exception applies, for instance for hospitals or political parties.[64]

iv)    Informed Consent

Moreover, installing a browser plug-in qualifies as *"the storing of information, or the gaining of access to information already stored, in the terminal equipment of a subscriber or user"* to which additional rules apply. Under the e-Privacy Directive, storing or accessing information on a user's device is only allowed after the individual's informed consent.[65]

Thus, the only available legal basis to collect data about an individual's browsing activity, and to further process those data, is to obtain a participant's "explicit consent." The requirements for valid consent are rather detailed and strict in Europe. Consent requires a *"freely given specific and informed indication of his wishes by which the data subject signifies his agreement to personal data relating to him being processed."*[66] In brief, valid consent cannot be obtained through the fine print in terms and conditions, and tacit consent (for instance with an opt-out system) is not sufficient.[67] In other words, research participants will need to be informed explicitly and extensively about the data types we collect, for which purpose, what their rights are regarding that data, etc.

Informed consent is likely the only viable legitimate ground to base our research on. However, this approach is far from ideal: a consent form that is too detailed may confuse panelists, overburden them with difficult information, and may

---

be fulfilled by further processing of data which does not permit or not any longer permit the identification of data subjects these purposes shall be fulfilled in this manner.").

[63]    *See* Data Protection Directive art. 7.

[64]    *Id.*, art. 8.

[65]    Directive 2002/58/EC of the European Parliament and of the Council of 12 July 2002 Concerning the Processing of Personal Data and the Protection of Privacy in the Electronic Communications Sector (Directive on Privacy and Electronic Communications) 2002 O.J. (L 201) 1, https://perma.cc/C54R-XVKB art. 5(3).

[66]    Data Protection Directive art. 2(h).

[67]    *See* Article 29 Data Protection Working Party, Opinion 15/2011 on the definition of consent, (July 13, 2011), https://perma.cc/H6CC-7Y4D.

even scare them away. Moreover, there is a growing body scholarship that questions of the usefulness and effectiveness of consent requests.[68]

v)     Security

Finally, data protection law provides that data controllers, such as a university, must secure the data appropriately.[69] Often, the available research infrastructure at universities is not fully in tune with the law's security requirements, meaning additional measures must be taken to ensure that the transmission and storage of personal data in the course of scientific research complies with the law.

b)     <u>Complying with Research Ethics</u>

The legal requirements mentioned above are not the only constraints, however. Research ethics add extra layers of complexity to the problem. Research and educational organizations, their organizational units, and funders (such as the EU's H2020 program[70]) all have ethical guidelines, review procedures and boards, and institutional and procedural safeguards. It is important to note, though, that empirical research in legal scholarship has less history than, for example, the social sciences, and thus the research ethics review infrastructure process is still nascent. Therefore, tensions can arise at levels as basic as attribution and the qualification of authorship or the acceptability of different sources of funding, but also at more organizational (e.g., different oversight committees and cultures) as well as substantive levels (e.g., potential conflicts between research ethical requirements and data protection law).

An interdisciplinary project such as ours that integrates communication science with legal research raises particular issues for ethics review. Currently, every social scientific data collection effort carried out at the University of Amsterdam is overseen by an ethics board. In addition, the project must pass the ethical review board of the EU research funding body, which conducts its own investigations into ethical matters.

---

[68]     *See, e.g.,* FREDERIK J. ZUIDERVEEN BORGESIUS, IMPROVING PRIVACY PROTECTION IN THE AREA OF BEHAVIOURAL TARGETING (2015); Solon Barocas & Helen Nissenbaum, *On Notice: The Trouble with Notice and Consent,* PROC. OF THE ENGAGING DATA FORUM: THE FIRST INT'L FORUM ON THE APPLICATION AND MGMT. OF PERSONAL ELECTRONIC INFO. (Oct. 2009); Daniel J. Solove, *Privacy Self-Management and the Consent Dilemma,* 126 HARV. L. REV. 1879 (2013); Alessandro Acquisti & Jens Grossklags, *What Can Behavioral Economics Teach Us About Privacy? in* DIGITAL PRIVACY: THEORY, TECHNOLOGIES AND PRACTICES 363 (Alessandro Acquisti, et al. eds, 2007).

[69]     *See* Data Protection Directive art. 17(1).

[70]     The H2020 Framework of the European Union is a large-scale research program, funding basic and applied research in every scientific domain. *See Horizon 2020: The EU Framework Programme for Research and Innovation,* EUROPEAN COMMISSION, https://perma.cc/WX6P-8CXK.

The core principles of these different review boards are essentially the same. However, important differences remain, as the different boards have different priorities, concerns, past experience, and approach to the practical dilemmas. For example, research that may seem unfamiliar (or even intrusive) in the realm of legal research might be considered by a social science or medical ethics board to be standard practice, and accordingly judged along differing standards.

The lack of a coherent research ethics review infrastructure creates particular obstacles for interdisciplinary projects. Indeed, some of the ethical considerations are inherently in tension with each other, and with other, legal principles, such as data protection law. For example, under data protection law researchers are charged with strictly controlling and limiting the sharing of personal data with others. In contrast, research ethics often encourage researchers to share research data widely as 'open data'.[71] Thus, norms of scientific integrity and transparency can conflict with the legally and ethically necessary protection of the privacy of the respondents.

To illustrate: the ethical guidelines of the Amsterdam School of Communication Research state that:

> The data that are gathered in the course of the research are not passed on to third parties (neither published nor disclosed in conversations or mutual consultation) in such a way that allows the results or other findings to be traced back to individual test subjects. An exception to this is research in which the results of earlier research are presented as a criterion for selecting test subjects. In this case, the data are encrypted as securely as possible when exchanged, and they are never disclosed to anyone other than the individuals involved in conducting the research. Of course, in such cases the data are anonymized after they are collected, and the resulting publications and suchlike always use anonymous data.[72]

---

[71] *See generally* Zuiderveen Borgesius, Van Eechoud, and Gray, *supra* note 57.

[72] Department of Communication Science University of Amsterdam Ethics Committee, *Ethical Review for Research at the Department of Communication Science,* UNIVERSITY OF AMSTERDAM (Sept. 5, 2013), https://perma.cc/6MX7-86RG.

In line with the legal requirements, researchers are responsible for securing the privacy of the test subjects under all circumstances. But pseudonymization does not provide sufficient protection of their privacy. Hence, to comply with EU law, we must take additional measures to safeguard the privacy of our research participants.[73]

Yet, to take these additional steps to ensure the privacy of the respondents could conflict with norms of scientific integrity and transparency. Scientists are told to be as transparent and verifiable as possible regarding data collection and data analysis. This line of thought suggests that all raw data should be stored in a repository accessible for peer reviewers who want to check the validity and reliability of the research. The current open access approach to science further advocates that research data should be made openly accessible. For example, the Netherlands Code of Conduct for Scientific Practice of the Association of Universities in the Netherlands (VSNU) expressively highlights the aspect of verifiability of results and data. It requires not only that 'raw research data are stored for at least five years.'[74]. The code also demands that the raw data are made available to other scientific practitioners at request, and done so in a way 'that they can be consulted at a minimum expense of time and effort'[75] – requirements that contradict the demands from data protection law to minimize and restrict the sharing of raw, non-anonymized data with third parties.

Therefore, researchers on the one hand face an increasing pressure to adhere to high transparency standards, while on the other hand, face ever-stronger calls to strengthen privacy protection and limit public data sharing that could violate the privacy of participants.[76]

c)        Other legal regimes

In the above analysis we have focused primarily on the challenges for research that arise from privacy and data protection regulation. Although it would exceed the scope of this publication to analyze them in greater depth, many other areas pose issues for researchers investigating algorithms, and for our research in particular. One such area is intellectual property protection. Part of our research design is to collect data from websites, such as news websites or social network websites. Under certain conditions, these websites can be subject to additional protections under intellectual property law. One example would be the so-called database right, which protects substantial investment that has been made into the collection of data.[77] The decision of whether database rights apply to a website, which would permit the website owner to prohibit scraping and copying content from the website, is subject to a case-by-case evaluation.[78] Database rights, however, also provide for a research exception.[79] In Europe, a new copyright law exemption for text and datamining for scientific research has recently been proposed.[80]

---

[73]    Recent research in the field of market research has come to similar conclusions, *see, e.g.* Daniel Nunan & MariaLaura Di Domenico, *Market Research and the Ethics of Big Data*, 55 INT. J. MARK. RES. 2–13 (2013).

[74]    *The Netherlands Code of Conduct for Scientific Practice*, ASSOCIATION OF UNIVERSITIES IN THE NETHERLANDS 7 (2012), https://perma.cc/QFV3-AGLF.

[75]    *Id.*

[76]    For a recent review of securing the privacy of participants in United States medical research, see Bradley Malin, David Karp & Richard H. Scheuermann, *Technical and Policy Approaches to Balancing Patient Privacy and Data Sharing in Clinical and Translational Research*, 58 J. INVESTIG. MED. 11–18 (2010). On the balance between data privacy and open data, see Zuiderveen Borgesius, Van Eechoud, and Gray, *supra* note 57.

Another recent development that has gone widely unnoticed by the European algorithm research community is the adoption of a new directive that fortifies the protection of trade secrets in Europe.[81] The directive defines trade secrets as "information which . . . is secret in the sense that it is not . . . generally known among or readily accessible to persons within the circles that normally deal with the kind of information in question; . . . has commercial value because it is secret; . . . [and] has been subject to reasonable steps under the circumstances, by the person lawfully in control of the information, to keep it secret."[82] Arguably, algorithms can fall under that definition. If they do, the holder of a trade secret can prohibit any unauthorized access to, appropriation of, or copying of the materials from which the secret can be deduced.[83] Interestingly, the directive provides for an exemption for journalists, but not academics.[84]

---

[77] *See* Directive 96/9/EC of the European Parliament and the Council of 11 March 1996 on the Legal Protection of Databases 1996 O.J. (L 77), 27.3.

[78] The authors wish to thank Marco Caspers for pointing this out. For further details on the situation under which database rights could apply, see Triaille, Jean-Paul, Jérôme de Meeûs D'Argenteuil, & Amélie de Francquen, *Study on the Legal Framework of Text and Data Mining (TDM)*, DE WOLF & PARTNERS (March, 2014), https://perma.cc/FXF3-RNWC; Hargreaves, Ian, Lucie Guibault, Christian Handke, Peggy Valcke, Bertin Martens, Ros Lynch, *et. al*, *Standardisation in the Area of Innovation and Technological Development, Notably in the Field of Text and Data Mining: Report from the Expert Group*, EUROPEAN COMMISSION – DIRECTORATE-GENERAL FOR RESEARCH AND INNOVATION (2014), https://perma.cc/3BA2-AXQZ.

[79] *See supra* note 77, art. 6. *See also*, Triaille, de Meeus & Francquen, *supra* note 78, at 79-80.

[80] *See Proposal for a Directive of the European Parliament and of The Council on Copyright in the Digital Single Market*, art. 3, COM (2016) 0593 final (Sept. 14, 2016), https://perma.cc/5YN7-BPNF.

[81] *See* Directive (EU) 2016/943 of the European Parliament and of the Council of 8 June 2016 on the Protection of Undisclosed Know-How and Business Information (Trade Secrets) Against Their Unlawful Acquisition, use and disclosure (Text with EEA relevance), O.J. (L 157), 15.6.

[82] *Id.* art. 2 (1).

Finally, is important to realize that website owners also have the means to regulate what is permitted or not permitted on their websites, in the form of community guidelines and terms of use. This creates an additional hurdle. Some sites are more permissive of using parts of the side for research (e.g. Twitter)[85] than others (e.g. Facebook).[86] The varied overall picture creates additional legal uncertainty for researchers. Another source of uncertainty is whether and under which conditions researchers are bound to the terms of use at all.[87]

### 4.    *Panel constraints and organizational limitations*

We also have to consider how certain organizational limitations and practical realities may affect the study results. As we discussed earlier, one of the key components in a distributed/collaborative audit approach is having access to the right mix and number of respondents, who are willing to share their browsing behavior. It is obvious that in this case the usual convenience samples and self-managed approaches[88] will be insufficient. Therefore, we cooperate with a professional research organization, which manages a large enough panel that is representative of the Dutch population along key sociodemographic variables, such as age, education, income, etc. The advantages of working with a professional research panel are obvious: respondents are used to being surveyed, there is a plethora of data available on them from previous surveys, the research organization fulfills important tasks, such as recruitment, panel management, design and communication, etc. These advantages, however, come at the additional cost of needing to account for the considerations of both the panel participants and the interests of the research organization.

---

[83]    *Id.* art. 4 (2)(a).

[84]    *Id.* art. 5.

[85]    *Twitter Terms of Service*, Twitter, https://perma.cc/3Z4H-RFWV (last visited January 22, 2017).

[86]    *Statement of Rights and Responsibilities*, Facebook, https://perma.cc/54MC-V8NE  (last visited January 22, 2017).

[87]    *See e.g.* Triaille, de Meeus and Francquen, *supra* note 78 at 73-74.

[88]    Many social science studies rely on "convenience samples" -- in other words respondents who are cheap and easy to access, such as college students, respondents recruited on the internet, or paid for via online collaboration platforms, such as Amazon Turk. While such samples are cheap, and require no third party to manage the respondents, the reliability of these studies are severely limited by the lack of representativity, missing (sociodemographic, etc.) information, and other, known and unknown biases inherent in working with such groups.

As previously discussed, securing informed consent from our panelists is not simply an organizational requirement. Informing our participants clearly and comprehensively about what we plan to observe, what kind of sensitive personal information we collect, and how we secure and store their data is both a legal and ethical obligation. But this obligation has far-reaching consequences on the practical level.

First, we face the issue of how to explain highly complex technical, legal, and ethical issues in simple and straightforward terms so it remains accessible for the average Dutch internet user. However, the more clearly we explain what we do, the lower the participation rate might be, especially among the more privacy-conscious respondents. If privacy-sensitive individuals do not join the research with the plug-in, our sample group may be biased. While our legal and ethical obligations dictate that panel participants be constantly aware if and when they are under observation, this awareness is unfortunately less than desirable from a research perspective, where we expect people to behave more naturally when they forget that they are being observed.

Most companies solve this problem of informed consent by burying the more controversial clauses in the fine print of the Terms of Service. Such an approach breaches data protection law,[89] and is not an option in our case. Instead, we have few other alternatives than to be as transparent, intelligible, and comprehensive as possible about our goals, methods, and safeguards. Various tools are employed to achieve this: a well-designed consent form, and radical transparency in our interactions with participants.

Organizational considerations can also affect the outcome of the study. Our research partner organization that manages the panel expects its reputational and institutional considerations to be taken seriously. The organization's concerns are partly overlapping with ours, in terms of reputation, data security, legal compliance, etc. In addition, the company rightly wants to prevent anything that might negatively affect their long-term investment in their panel, by, for example inducing churn, higher non-response rates, loss of trust, etc. Having strong incentives to err on the side of caution, their organizational concerns oblige us to impose an extra layer of personal data filters, which remove information that we can legally collect, but may identify respondents, such as home addresses, or email addresses. .

---

[89] *See* Article 29 Data Protection Working Party, Opinion 15/2011 on the definition of consent, *supra* note 67 at 35 ("The information must be provided directly to individuals. It is not enough for it to be merely available somewhere").

## V.        BALANCING INDIVIDUAL RIGHTS AND SOCIETAL NEEDS IN RESEARCH INTO ALGORITHMIC AGENTS: SOME LESSONS LEARNED

As the previous Part demonstrated, the process of designing a research regime for "monitoring the algorithms" is extraordinarily fraught with practical, legal and ethical dilemmas. The legal principle of data minimization conflicts with a researcher's desire to collect massive amounts of data. The purpose limitation principle makes it harder for a researcher to remain flexible in a quickly changing environment, where new, unexpected developments are expected to appear at any moment. Data protection law's rules about data retention, security, and safety may conflict with research ethics on transparency and accountability. And the obligations to acquire informed consent may compromise some research objectives, even when informing users has limited effect as a privacy protection measure.[90]

In the following section, we develop some suggestions on how to solve the legal and ethical challenges discussed in the previous Section. We suggest in the following pages that addressing the legal and ethical dilemmas requires a multi-tiered approach that combines transparency with technical and organizational measures.    It is our hope that these recommendations contribute to a growing conversation on ways to conduct responsible research into algorithms and personalized experience cocoons.

### A.      *Designing ethical research: transparency*

Transparency and informed consent have long been regarded as the main lines of defense in privacy protection. However, there is growing criticism of this approach. For instance, even when an organization fully discloses how it uses personal data, individuals may not be able to understand all the information provided to them. Individuals may not foresee the possible consequences of disclosing data. And even if individuals

---

[90]    *See* ZUIDERVEEN BORGESIUS, *supra* note 68., in particular chapter 7 (p. 187-222)..

understood all the information an organization provided, they might not act upon it.[91]

Nevertheless, transparency should be a key element in efforts to minimize risks and harms associated with research. Being transparent about what data researchers collect – and to what end  - is an expression of the respect for the autonomy of the user, and thereby also an ethical requirement. But transparency is also an element of accountability. By providing specific information about processes and conduct, users and third parties will be able to measure actors against what they promised to do or not to do. And being transparent also forces those collecting and processing the data to think carefully about the "whys" as well as the "whats". We suggest some guiding principles regarding transparency below.

### 1.     Informed consent and transparency

As noted, under EU law in "informed consent" is likely the only viable legitimate legal basis to guide our research, regardless of scholarly criticism of consent as a privacy protection measure. Hence, we suggest that the form and content of the consent form should be of crucial importance in research design.

The preparation of the documents which inform research participants about the depth and scope of our data collection activities, and which asks them to consent to these activities, gave us the opportunity to reuse our insights from research into transparency in general, and informed consent in particular.[92] The legally (and ethically) appropriate form of acquiring informed consent may, however, conflict with other aspects of the research. A consent request that is too detailed may confuse or even scare panelists. For example, based on the feedback from our research partner in charge of the panel about the length and complexity of our initial consent form, we were forced to rethink the presentation and wording of the text, without compromising on the content.[93]

---

[91]     *See generally* Alessandro Acquisti & Jens Grossklags, *What can behavioral economics teach us about privacy*, *in* DIGITAL PRIVACY: THEORY, TECHNOLOGIES AND PRACTICES 363–77 (Alessandro Acquisti et al. eds., 2007); ZUIDERVEEN BORGESIUS, *supra* note 68.; Natali Helberger, *Form Matters: Informing Consumers Effectively*, SSRN (2013), http://ssrn.com/abstract=2351791.

[92]     This is an example where legal research can be relevant not only for academics, policy makers, and stakeholders, but can also provide insights for the actual *process* of doing research itself.

[93]     Designing a revised and better structured consent form is no trivial task, and one that took up more research time than anticipated. Part of that effort was

We developed a more general "privacy notice" for our website that explains what the project is about, which research questions we look into, what kinds of data we collect, how we use the data, whether we share the data, and how we secure the data.[94] Sharing detailed information about the research with users is not only a matter of compliance.[95] Clear information is also a matter of respect for the user and a means to secure the users' active cooperation. Ideally, being open and clear about what we do will have the effect of winning the users' trust, which is indispensable for this project, and making them active participants to the research, rather than merely research subjects.

## 2.    *Data retention and transparency*

It is difficult to find the right length of time for the data retention period. Data protection law's storage limitation principle requires that data be deleted as soon as possible if they are no longer necessary.[96] On the other hand, scientific and ethical norms increasingly dictate that underlying data are shared to enable review and reuse.

Similar conflicts currently play out on an ad-hoc basis, and they point to a need for a more general, systematic approach to resolving conflicting ethical goals. Having said that, there is a long tradition of research that involves sensitive data, and there are alternatives to full access to the data that safeguard both the privacy of the participants and transparency of the research conducted. One such alternative is that the access to highly sensitive raw data be safeguarded by an ethics committee.[97] In practice, this means that if during the peer review process a reviewer requires access to the data in order to evaluate whether the results have been obtained correctly, the reviewer contacts the ethics committee which ensures that the data are not shared with any other parties and only the data required for the request are made available. Research institutions should explore the feasibility of similar solutions.

## 3.    *Transparency as encouraging dialogue*

---

also to cooperate together with an UX designer, and experiment with alternative means of informing potential participants, such as using video. These are research costs that are typically not accounted for in initial budget calculations, but the efforts needed to comply with technical and ethical standards certainly deserve a place in the budgets for research proposals.

[94]   See the privacy notice at https://robin.personalised-communication.net/ .

[95]   *See* Data Protection Directive art. 10.

[96]   *See* Part IV.B.2.a.ii, *supra*.

[97]   See, for example, the procedure recommended by academic publisher PLOS | ONE (specialized in medical research), available at https://perma.cc/GU3Q-HWK7..

Finally, the public discussion about the use of data analytics, algorithms, and their influence on fundamental rights and values such as privacy, freedom of expression, personal autonomy, or the right to non-discrimination is still very much in flux. Often, it is not only companies, regulators, or governments that do not know the right way to discuss these deep issues, but scholars as well. Transparency has a role to play in this debate. One way of dealing with this dynamic is to make the choices, conflicts, and possible solutions as clear and accessible as possible. This is a goal of this paper as well: to cultivate scholarly debate about how to research the black box.

## B.    *Deploying the principles in practice*

A commitment to transparency may be key, but for reasons explained above, informing users alone is likely an inadequate response to the challenges that research into the black box can pose. This is why we also implemented a number of technical solutions to reduce the scope and depth of our observation to a justifiable minimum, and made reasonable efforts to discard information that is highly personal, sensitive, or non-essential for our study.

### 1.    *Whitelists*

We decided to limit the amount of captured data by restricting our observations to a relatively small set of websites, which are relevant to our original topics of personalization in news, commerce and health. For this reason the browser plugin routes the browser traffic through the transparent proxy only if the visited website is on a pre-defined whitelist.

To include a website in the whitelist, researchers should provide a detailed description,[98] which provides not just the practical and theoretical justifications for inclusion, but also the risks of collecting unwanted, sensitive, personal information from that website. These inclusion requests should then be approved by the joint ethical review board, described in the next Section.

Providing justification and explanation for the inclusion of a website in the whitelist has four benefits. First, a justification for the observed websites helps to comply with the legal obligation of data minimization. Second, justifying inclusion on the whitelist forces us to precisely define why we want to observe user interaction on each particular website. Third, it allows us to inform participants about what we do and why, as participants are able to consult the explanations online. Finally, we hope that through this public information, users are more willing to trust us as we communicate that we do not collect information indiscriminately, and that we avoid observing particularly sensitive websites such as the websites of banks, doctors, and other highly-personal pages.

---

[98]    See Annex 1: Justification form for website whitelisting.

We organize the websites on the whitelists into specific categories, such as Dutch language news websites, Dutch price comparison websites, and international news websites.[99] This category-based approach is important for several reasons. As we expected several hundred websites to be whitelisted, such categories help us to present this overwhelming amount of information in a user-friendly manner, which hopefully translates into well-informed users and contributes to high consent rates[100]. By defining such categories, and by limiting the number of categories and the number of URLs in each category, we also aim to further comply with the data minimization principle.

The whitelist approach comes with a number of compromises, however. Restricting the scope of our observation to a limited number of websites no doubt affects the effectiveness of research.[101] The whitelists also introduces a certain amount of inflexibility in the work process. It is extremely difficult to predict which websites should be included in a whitelist. While the periodic update of the whitelist is possible, each update to the whitelist might require the renewal of the consent forms. Finally, the entire process of designing the whitelists was very time and resource intensive.

2.    *Filters*

Data collected from whitelisted websites might still contain information that we would like to avoid capturing, or that has no relevance to our research objectives. To filter out things like private correspondence, financial information, bank account data, passwords, etc., we maintain a set of filters that aim to remove such information from the captured data stream before it reaches storage.

---

[99]   See Annex 2: Whitelisted website categories for examples.

[100]  In a preliminary study on the informed consent process we measured a 50% consent rate. During the live recruitment the actual rate was slightly lower.

[101]  This will be more true for some areas (e.g. commerce) than for others (politics) because the differences between the way profiling and targeting for behavioral advertising or in the news media works.

Despite our best efforts, the filter approach has limitations. As filters operate on the body of data as defined by the individual websites on the whitelist, any change to how a particular website transfers a piece of information to be filtered requires an update to the filter in question. This requires the continuous maintenance of a large amount of filters, which are prone to become obsolete at any subtle change in the website without warning. Hence, filtering out *all* unwanted information is not guaranteed. Nevertheless, the filters do help to minimize the collection of unnecessary data.

### 3.    *Secure storage*

Security is another important and challenging issue. Data protection law and common sense requires us to keep user data safe and secure. Ensuring data security can be difficult as universities typically lack the technical infrastructure as well as the expertise to provide industry-standard data security.

Our strategy to achieve reasonable levels of data security had several components. First, we decided to involve the Dutch national research computing infrastructure provider, "SURF".[102] Since they provide data storage and processing services to several other privacy sensitive research projects, such as health and genomics research, they have the expertise we require. In the light of the current concerns about data transfers, the strict requirements about data security in data protection law, and the recent judgment of the European Court of Justice invalidating the Safe Harbor for US-EU data transfers, it is also important that the data is stored and processed in an EU member state.[103] On top of the secure infrastructure, we use strong encryption technology every time data is stored or transferred.

### C.    *Creating organizational safeguards*

The third part of our strategy consists of a number of organizational measures. First, we entered the obligatory data processing agreements with our partner organizations.[104] Beyond these minimum legal requirements we decided to closely involve our partner organizations in the project governance. We set up two bodies to deal with unforeseen issues: a working group on privacy and ethics, and a privacy steering committee.

---

[102] SURF is the collaborative ICT organization for Dutch higher education and research. It offers scientists in the Netherlands access to state-of-the-art computing facilities.

[103] *See* Case C-362/14, Schrems v. Data Protection Comm'r (2015). For commentary, see Christopher Kuner, *Reality and Illusion in EU Data Transfer Regulation Post Schrems*, 18 GERMAN L. J. (forthcoming 2017).

[104] *See* Data Protection Directive art. 17(3).

The working group on privacy and ethics consists of a number of members of the project team. Their main task is to ensure compliance of the Personalized Communications Project with legal and ethical requirements. In addition, the working group is in charge of the operational aspects of legal compliance, including managing the consent declaration and its documentation; monitoring ongoing compliance of personal data handling; responding to complaints and react to other information; handling third party requests for access to the datasets and communicating decisions to the research team and the research partner. By making compliance a separate management task and dedicating personnel-power to this, we aim to guarantee continuous attention for matters of privacy and data protection within the project.

In addition to the working group, we set up the privacy steering committee, with an equal number of members from CentERdata (our research partner), the University of Amsterdam, and an external member familiar with issues of privacy and research ethics. This body acts as a joint ethics advisory board, with extensive veto powers over all of the issues that involve the research panel, including such key components of the research as the form and content of the consent declaration and privacy notice; the technical specifications of the observation infrastructure; key features of the whitelisted websites, such as the generic categories; the data management policy of the collected data; and possible complaints by the survey participants.

We devised these technical, organizational and procedural frameworks in response to the concerns we identified during the design of our research methodology. It should be noted that the most effective frameworks will be in a constant state of flux, as new challenges force actors to provide adequate responses. Nevertheless, these safeguards offer a reliable framework to address unforeseen problems.

## VI.        CONCLUSION: TOWARDS A VISION ON RESPONSIBLE BIG DATA RESEARCH

In the dawn of the information age Isaac Asimov, the science fiction writer, formulated three law of robotics.[105] Half a

---

[105] (1) A robot may not injure a human being or, through inaction, allow a human being to come to harm. (2) A robot must obey orders given it by human beings except where such orders would conflict with the First Law. (3) A robot must

century later, having autonomous artificial agents around us is no longer science fiction. Despite the ubiquity of these agents, we are yet to have anything remotely similar to Asimov's three simple laws. The main reason for that is that autonomous algorithmic agents are not what we imagined them to be. They are not the huge, shiny, metal robots, or standalone machines with superhuman strength who have the potential to cause physical harm to humans in various local contexts. Instead, algorithmic agents operate on an immaterial level; they are networked and interconnected; and rather than being engaged in local, one-on-one interactions with individuals, they operate on planetary scale, simultaneously affecting the lives of billions. Despite, or, perhaps because of, these differences, our societies may still need rules to avoid, or at least minimize, potential harm from our algorithmic agents. Yet we still have no idea how to translate Asimov's simple and straightforward instructions into something applicable to our current context.

To arrive at solutions, we first need to be able to answer some fundamental questions on how algorithms and humans co-shape society. How do algorithmic agents improve our individual and communal lives, and how do we encourage such developments? What are the possible injuries – through their design, operation, decisions, or their negligence – that algorithmic agents can cause to human individuals and communities? What kind of values do we encode into algorithmic agents? Who encodes those values, and at which points? How do we, as a society, teach our values to algorithmic agents, and how do those get reflected back to us? How do we detect if something is going astray? Do researchers have a public obligation to do so? How do we assign individual and communal responsibilities and accountabilities in this domain? How do we account for the plurality of often contradicting values that algorithmic agents need to comply with? What kind of informal practices, norms, professional ethics and codes of conducts - and what kinds of legal instruments - are developing around the existence of algorithmic agents? Where is this piecemeal, bottom-up regulatory scaffolding lacking? Where does it need top-down, planned reinforcement?

To answer any of these questions, first we need to know what is actually happening as humans and algorithmic agents interact. We need to be able to observe these interactions as they take place; to be able to gather empirical evidence on those interactions so we can start exploring their effects. Without this

---

protect its own existence as long as such protection does not conflict with the First or Second Law. ISAAC ASIMOV, I, ROBOT 37 (1950).

knowledge, any answer to the questions above must remain speculative.

Yet, the acquisition of this insight into human-algorithmic agent interactions poses serious challenges and paradoxes. In order to fully understand the effects and implications of algorithmic profiling and targeting, it is necessary to engage in the collection of large amounts of data, and thereby behave similarly to the companies whose behavior and impact we research. This means that researchers may be complicit in any intrusion into the privacy and personal data of their research subjects. Should ethical norms and legal requirements render such research impossible? And should different conditions apply for research that is being done behind closed doors at commercial companies, or by academic researchers, who operate under elaborate ethical demands of transparency and accountability?[106]

Doing research and advancing knowledge and science are not goals that, in themselves, justify violating the rules that were meant to protect privacy and related rights, or violating the values that are the subject of our research. But without the right tools, we as a community of scientists, citizens, and humans lose our ability to reflect on one of the most important developments in the history of communication. And seeing the complexity of algorithms and the challenge of understanding their effects on users, society, and the values that our societies hold, it can be argued that academic researchers, as (hopefully) independent and skilled observers, have an important role in advancing knowledge and understanding.[107]

For research which relies on tools that gather large quantities of potentially highly sensitive data on a relatively large group of participants, being aware of the conflicts and the stakes at hand is the first step towards developing more responsible methodologies. The next step is to devise solutions that make the use of these tools safe. While most of the solutions will be different from project to project, we hope that this contribution offers some more generalizable approaches as well.

We believe that one ideal long term solution would be to set up representative national panels which would allow full insight into their members' online behavior for research purposes. As, for example, Nielsen tracks TV viewing, similar institutions, both public and/or private, would enable insights into how a representative sample of individual internet users interact with various online services.[108] We believe that such oversight, enabled and overseen by independent third parties, is essential to gather knowledge on the developments of our societies.

---

[106] The GDPR has a very broad research exception. *See* Part IV.B.3.c., *supra.*

[107] Or as the GDPR has framed it: "For scientific or historical research purposes or statistical purposes, the legitimate expectations of society for an increase of knowledge should be taken into consideration." General Data Protection Regulation, recital 113.

But lacking such a public "looking glass" infrastructure at the present, we have tried to emulate something similar, as closely as the current legal and ethical constraints allowed us. We have to rely on an informed consent approach, seeking individuals who are willing to participate in our research for their own, individual reasons. But rather than trying to exploit the fact that individuals are susceptible to consent to anything, without giving a serious second thought to what they click to agree on, we decided to take a more difficult path. This path relies on being fully transparent about both the scope and depth of our activities, in combination with technical and organizational safeguards. But securing informed consent and doing so in a way that is actually meaningful is only one part of the solution. The other part of the solution is to build ethical and normative safeguards into the actual research environment. This approach, and the subsequent system of technological, ethical, and organizational safeguards we have built, may serve as an example that others can also rely on.

Our approach comes at a cost: by informing panelists properly about our data collection, researchers risk losing their trust. By limiting our tool to a limited number of websites on a whitelist, we not only forgo some of the technical potential but probably also limited the usefulness and validity of our research. And we have to re-purpose significant amounts of research time to designing the data management strategies, justification forms, governance structures, and the like. Funders should provide adequate financial resources to make research legally compliant, and ethically responsible. And society must decide under what conditions the public and scientific interest in more algorithmic transparency justifies conflict, for example with restrictive terms of use of (commercial) exploiters of algorithms but also intellectual property law and data protection.

This brings us to more fundamental issues. One is the question of when the potential (monetary, but also social and individual) costs of such a monitoring infrastructure outweigh its potential positive contribution to science. This is a difficult question to answer because both costs and benefits are difficult

---

[108] *See* NIELSEN, https://perma.cc/FZ5N-BVE2. While Nielsen is a commercial entity, offering its metrics on the market for a fee, the research panel we are collaborating with is essentially financed by the Dutch public, and thus access to it is much cheaper and easier, and the results of the research are more easily accessible.

to quantify, and to do so would require even more research along similar terms.

Possible lessons could be learned from the principle of proportionality that applies, for instance, to governments, but also in the data protection context. A leading principle for the activities of governments is that where there are lighter, potentially less invasive alternatives, those should be chosen. For researchers this could mean that when choosing between different methods to research algorithms, their potential invasiveness and strain on research subjects needs to be an important consideration, in addition to the usual inquiries of potential effectiveness in answering a particular research question.

Another issue is whether there is a need for society to agree on some principles of responsible research into the algorithmic society. This question is perhaps controversial because such principles could interfere with academic freedom and the role that academics play in a democratic society. And yet, because of their societal role, academics have an equally important task in bringing light into matters that would otherwise not easily be exposed to the public. Formulating such principles could be a method ensuring that academics engage in research in an ethical, publically accountable way. Under the General Data Protection Regulation, using lawfully collected personal data in scientific research is foreseen if this adheres to recognized ethical standards for scientific research, in other words principles for academic research. Research guidance could take the form of broadly accepted (and ideally dynamic) standards, e.g. developed by researchers, in cooperation with data protection authorities, judges, ethicists, etc.[109]

In addition, there could be a role for governments to clarify and improve research exceptions in data protection, intellectual property, and contract law. Exceptions should account for the fact that doing such research can often be in the public interest. A clearer distinction should be made between academic, publicly accessible research (research that is contributing to ending the algorithmic control crisis) and research that is taking place within companies, and behind closed doors (potentially further contributing to the control crisis). Lawmakers should also more clearly define the conditions under which organizational and technical safeguards are adequate and sufficient.

---

[109] For example, in the Netherlands, the leading universities in the area of data science research have formed a multi-disciplinary coalition consisting of computer scientists, legal scholars, ethicists, economists, communications scientists, psychologists, etc. to develop together principles and best practices of responsible data science research, the Responsible Data Science (RDS) Consorium. *See* RESPONSIBLE DATA SCIENCE CONSORTIUM, https://perma.cc/XG9A-78MU.

Data protection law, the General Data Protection Regulation, and even copyright law currently acknowledge the societal role of research in algorithms. Well-balanced research exceptions play an important role in this. But we have also identified instances in which the requirements of research in algorithms have been ignored, such as in the case of trade secrecy protection in Europe, or the terms of use of the very companies that use the algorithms that researchers try to investigate. This is a problem not easily solved. It is true that commercial companies are investing significant amounts of money in technology development and at the same time benefiting from the fundamental freedom to conduct business and protect their property. It is also true that some of these technologies are likely to have a significant impact on society, and society still needs to learn what these impacts are. Algorithms are an example of this. Should it be possible that research into algorithms is legally impossible because of terms of use? Does such a situation create information asymmetries that may not only affect individual users, but society at large? If so, should there be limits to contractual freedom not only in the interest of individual users, but society and academics to the extent that they do research to advance transparency? These are relevant questions, and we suggest that researchers continue puzzling about them.

As a final point: ways of conducting responsible research into algorithmic society should be acknowledged as a research topic in itself. Much of the research into algorithms, but also into data protection and privacy law, is directed at uncovering the ways in which algorithms can potentially comply or conflict with fundamental rights and values, such as privacy, non-discrimination, and freedom of expression. The results from this research should be used not only to improve laws and demand more societally acceptable algorithms. They should also be used to help researchers design better and more responsible research.

In other words, there is currently no ready-made recipe for doing research into algorithms and society. But we hope that our paper may contribute to solving the algorithmic control crisis by outlining the wider technical, legal, and methodological challenges that accompany attempts to systematically observe and aggregate the behavior of algorithmic agents.

## VII.    ANNEX 1: JUSTIFICATION FORM FOR WEBSITE WHITELISTING

### Draft form for websites included in the whitelist

| General | | | |
|---|---|---|---|
| URL | [Enter the URL of the website] Alias: [Enter the URL of an alias to the website] | | |
| Generic category | [Enter the relevant generic category within which the website shall be included on the whitelist] | No. | [count the website on the whitelist] |
| Reason for inclusion in the whitelist | [Insert a generic but precise reason for inclusion of the website on the whitelist in order to answer which research question] | | |

| Features of the website | |
|---|---|
| SSL | [Does the website or subparts of it use encryption and when, e.g. whether especially sensitive personal data is transferred, e.g. payment data] |
| Bi- or pluri-lateral commu-nications | [Does the website contain non-public, bi- and plurilateral communications, such as groups, messaging, chats?] |
| If yes | [Add URLs of pages where such communications takes place] |
| Includes data of third parties | [Does the website include personal data of individuals external to the participants who have agreed to the research? E.g. user-generated content.] |

| The following categories of personal data involved? | |
|---|---|
| user name/ password | [Yes/ no] |
| If yes | [Add URLs of pages where such data are involved] |
| sensitive personal data | [Yes/ no, e.g. health, political beliefs, sexual orientation] |
| If yes | [Add URLs of pages where such data are involved] |
| financial information | [Yes/ no] |
| If yes | [Add URLs of pages where such data are involved] |
| email addresses | [Yes/ no] |
| If yes | [Add URLs of pages where such data are involved] |
| other | [Could there be other especially sensitive personal data, e.g. psycho quizzes] |
| If yes | [Add URLs of pages where such data are involved] |

**Measures**
[Insert description of the measures to limit intrusions into privacy, protection of sensitive personal data, communications secrecy and rights of third parties]

**Approval for inclusion in the whitelist**

| | |
|---|---|
| Author | |
| Submitted to working group | |
| Approved by PI's | |
| Approved by CentERdata | |
| Inclusion in whitelist | |
| Information of survey participants | |

### VIII.    ANNEX 2: WHITELISTED WEBSITE CATEGORIES

| Category | Sub categories | Number of websites in category |
|---|---|---|
| Dutch and international news websites | national new providers regional news providers opinion sites national broadcasters regional broadcasters online-only news sites weather sites pay-per-article news providers English, French, German language news sites Flemish news sites Flemish broadcasters | 117 |
| Political parties and other entities | Dutch political parties | 10 |
| Health websites | general health information websites, 'Commercial' health websites | 32 |
| Blogs and discussion platforms | blog providers collaborative filtering websites online fora | 17 |
| Business-to-consumer web shops | consumer goods services auction sites travel sites | 39 |
| Price comparison websites | price comparison websites | 11 |
| Digital entertainment websites | music streaming online radio online video video-on-demand social video | 57 |
| Search engines | search engines | 16 |
| Reference works and political discussion boards | Reference works political discussion boards political mobilization platforms | 18 |

| | political                    information platforms | |
|---|---|---|
| Social media | Facebook<br>other social media | 6 |

## A.    *Sample justification*

**Category:**

Political parties and other entities

**Reason for inclusion in the whitelist:**

Political communication is increasingly tailor-made. Parties and candidates focus their efforts on convincing opinion leaders and undecided voters, those increasingly few upon whom the outcome of an election hinges. Two important internet-related developments prompted this change in political communication. First, on a practical level, there are more and more entities that are able to observe us and collect data on various aspects of our behavior (and thus our preferences), ranging from credit card analyses to personalized online services. A whole data broker industry has developed based on this information, selling highly detailed dossiers on the majority of individuals in developed societies.

Second, the proliferation of social media services offer better targeting opportunities. Our highly detailed online profiles on social media websites enable even better micro-targeting opportunities for anyone willing to pay for the opportunity.

Both of these developments promise huge payoffs for political campaigns. Political parties in the United States are using microtargeting to maximize the impact of campaign spending. Experts expect similar developments to unfold in Europe in the near future.

Despite the entry of political entities into the data domain, it is unclear how these developments affect the foundations of our democratic system. We lack basic information on how the personalization of political communication takes place: how political parties utilize dataveillance, how they communicate online with citizens, and/or to what extent and to what effect they personalize their efforts.

Hence, we would like to include the websites of political parties for three reasons

a)    To find out to what degree political parties use their own communication tools to target potential voters by showing them personalized information.

b)    To observe the breadth of information citizens receive in a political campaign online. Are they primarily informed through news sources that might serve them personalized limited information, or do they also get information from the political parties themselves?

c)    How changes in the news consumption of a user affect political engagement. For example, did users that started to use primarily personalized news media of a particular ideological leaning visit political party homepages more often?

**Necessary extra data protection measures:**

Personal data of third parties need to be filtered in the discussion and comment sections.